\documentclass[%
 twocolumn,amsmath,amssymb,aps,pra]{revtex4}
\usepackage{float}
\usepackage{appendix}
\usepackage{graphicx}
\usepackage[dvipsnames]{xcolor}
\usepackage{dcolumn}
\usepackage{bm}
\usepackage{color}
\usepackage{latexsym}
\usepackage{epstopdf}
\usepackage{color}
\usepackage[english]{babel}

\usepackage{amsfonts}
\usepackage{bm}
\usepackage{natbib}
\usepackage{bbold}
\usepackage{enumerate}
\definecolor{myblue}{rgb}{0.2,0.2,0.8}
\definecolor{myzard}{cmyk}{0,0,0.05,0}
\definecolor{mywhite}{rgb}{1,1,1}
\definecolor{mywhite}{rgb}{1,1,1}
\definecolor{myred}{rgb}{1,0.,0.3}
\usepackage[colorlinks=true,citecolor=myblue,linkcolor=myred]{hyperref}

\begin{document}


\title{Bridging the gap between topological non-Hermitian physics and open quantum systems}

\author{\'{A}lvaro G\'{o}mez-Le\'{o}n}
 \email{a.gomez.leon@csic.es}
\affiliation{Instituto de Física Fundamental IFF-CSIC, Calle Serrano 113b, Madrid 28006, Spain}

\author{Tom\'as Ramos}
\email{t.ramos.delrio@gmail.com}
\affiliation{Instituto de Física Fundamental IFF-CSIC, Calle Serrano 113b, Madrid 28006, Spain}

\author{Alejandro Gonz\'alez-Tudela}
\email{a.gonzalez.tudela@csic.es}
\affiliation{Instituto de Física Fundamental IFF-CSIC, Calle Serrano 113b, Madrid 28006, Spain}

\author{Diego Porras}
\email{diego.porras@csic.es}
\affiliation{Instituto de Física Fundamental IFF-CSIC, Calle Serrano 113b, Madrid 28006, Spain}

\date{\today}

\begin{abstract}
We relate topological properties of non-Hermitian systems and observables of quantum open systems by using the Keldysh path-integral method. 
We express Keldysh Green's functions in terms of effective non-Hermitian Hamiltonians that contain all the relevant topological information. 
We arrive at a frequency dependent topological index that is linked to the response of the system to perturbations at a given frequency.
We show how to detect a transition between different topological phases by measuring the response to local perturbations. Our formalism is exemplified in a 1D Hatano-Nelson model, highlighting the difference between the bosonic and fermionic cases.

\end{abstract}

\maketitle


{\it Introduction.---} Topological phases of matter were first discovered in electronic systems~\cite{QHE1,QHE2} and since then, their properties in equilibrium have been thoroughly studied~\cite{TI1,TI2,TI3}. During the last decade, the exploration of topological phases in non-equilibrium and dissipative systems has attracted great interest~\cite{Floquet-TI,TI-Dissipation}, triggered by the observation of topological effects in photonic lattices~\cite{TopologicalPhotonics,Photonic-TI,ReviewOptics-Topology} and quantum simulators~\cite{CiracReview2012,RevModPhysNori2014,Quantum-Simulators1} and, more lately, applications in sensing~\cite{Sensing1,ExceptionalPoints4,Sensing-McDonald2020,Sensing-Koch2021} and amplification~\cite{peano2016,PhysRevLett.122.143901,Top-Winding-finite,PhysRevA.103.033513}. 

Characterizing topology in out-of-equilibrium systems is complicated by the presence of intrinsic gain and loss mechanisms which must be included in the dynamics. Several approaches have been considered to partially tackle this problem. In Ref.~\cite{TI-Dissipation}, the notion of topology by dissipation was introduced, showing that an engineering of jump operators can lead to the dissipative preparation of topological states. More recently, a criterion to define topological invariants from density matrices was proposed~\cite{UlhmannPhase,UlhmannPhase2}, and an extension of topological band theory to include non-Hermitian matrices has been introduced~\cite{PhysRevX.8.031079,PhysRevX.9.041015,PhysRevLett.124.056802,PhysRevLett.124.240404,PhysRevLett.126.216405,Kunst2021}. While part of the phenomenology can be explained in terms of non-Hermitian effective Hamiltonians, this neglects quantum jumps and the quantum noise of the dissipative dynamics, and thus, cannot consistently describe the steady state nor the experimental observables of the system \cite{Clerk2018,Clerk}. Therefore, the field would benefit from a more complete characterization of dissipative topological phases, which accounts for gain and loss mechanisms in many-body open quantum systems, describes bosons and fermions on the same footing, and links the topological properties to measurable correlation functions.

We undertake this task by using the Keldysh path integral formalism~\cite{Sieberer_2016}, which we use to characterize non-trivial topological phases of quantum open lattices, and to establish a link between topological indices defined for non-Hermitian matrices and physical observables.
(i) We establish a link between Keldysh Green's functions of quantum open lattices with gain/loss terms, and non-Hermitian matrices.
(ii) We present a topological characterization of non-equilibrium Green's functions that allows us to define frequency-dependent topological indices. Our formalism relies on a mapping from non-Hermitian matrices into topological insulator Hamiltonians. Non-trivial topological phases correspond to directional amplification of excitations at a particular frequency.
(iii) Our work leads to a definition of topological phase transition in quantum open lattices. Surprisingly, such phase transition can occur in frequency space between regions with different values of the frequency-dependent topological index.
(iv) We show how topological properties can be tested by measuring the response of the system to perturbations. 
(v) We illustrate our results by studying the bosonic and fermionic realizations of the 1D Hatano-Nelson model, and highlight the role of particle statistics in our topological characterization. 

{\it Keldysh path integral.---} Consider a quantum system in a lattice with particles described by bosonic or fermionic operators, 
$\hat{\psi}_j$ and $\hat{\psi}_j^\dagger$. The dynamics can be described by the master equation for the density matrix operator $\hat{\rho}$~\cite{petruccione}:
\begin{eqnarray}
\frac{d \hat{\rho}}{dt}
=
-i
\left[\hat{H},\hat{\rho}\right]
&+&
\sum_{j,l} \gamma^{(d)}_{jl}
\left(\hat{\psi}_{j}\hat{\rho} \hat{\psi}_{l}^{\dagger}-\frac{1}{2}\left\{ \hat{\psi}_{l}^{\dagger}\hat{\psi}_{j},\hat{\rho}\right\} \right)
\nonumber \\
&+&
\sum_{j,l} \gamma^{(p)}_{jl}
\left(
\hat{\psi}_{j}^\dagger\hat{\rho} \hat{\psi}_{l} -
\frac{1}{2}
\left\{ \hat{\psi}_{l} \hat{\psi}_{j}^\dagger ,\hat{\rho}\right\}
\right),
\label{eq:Lindblad}
\end{eqnarray}
where $\hat{H}$ is the Hamiltonian and $\gamma^{(d)}$, $\gamma^{(p)}$ are matrices describing decay and gain processes, respectively. 

An alternative to the operator formalism is the Keldysh path-integral method~\cite{kamenev_2011}. There, a time slicing procedure and the insertion of coherent states leads to a set of fields ${\{}\psi_{j,\pm},\bar{\psi}_{j,\pm}{\}}$, being $\pm$ the Keldysh contour where the fields act. These fields are complex variables in the bosonic case and independent Grassman variables in the fermionic one (i.e., $\bar{\psi}_{j,\pm}=\psi^{\ast}_{j,\pm}$ in the bosonic case).  
From Eq.~\eqref{eq:Lindblad} one finds the Keldysh action~\cite{Sieberer_2016}:
\begin{equation}
    S = 
    \int_{-\infty}^{t_f} dt
    \left[ 
    \sum_j  \left(
    \bar{\psi}_{j,+} i \partial_{t}\psi_{j,+}
    -  \bar{\psi}_{j,-}i\partial_{t}\psi_{j,-} \right)
    -i\mathcal{L} \right],\label{eq:action1}
\end{equation}
defined in terms of the Lagrangian:
\begin{align}
\mathcal{L} =
& -i\left(H_{+}-H_{-}\right)\label{eq:Lagrangian} \\
& 
 +\sum_{j,l}
 \gamma^{(d)}_{jl}
 \left( \psi_{j,+} \bar{\psi}_{l,-}  
 -
 \frac{1}{2}
 \left(
 \bar{\psi}_{j,+} \psi_{l,+}
 + \bar{\psi}_{j,-} \psi_{l,-}
 \right)
 \right) 
\nonumber \\
&
+\sum_{j,l}
 \gamma^{(p)}_{jl}
 \left( \bar{\psi}_{j,+} \psi_{l,-}  
 -
 \frac{1}{2}
 \left(
 \psi_{j,+} \bar{\psi}_{l,+}
 + \psi_{j,-} \bar{\psi}_{l,-}
 \right)
 \right).
\nonumber
\end{align}
Here, $H_{\pm}$ results from $\hat{H}$ acting on the $\pm$ branch of the Keldysh contour. Remarkably, the action in Eq.~\eqref{eq:action1} has the same form, irrespective of whether it is for bosons or fermions~\cite{kamenev_2011,Tesis}. Also, notice that gain $\gamma^{\left(p\right)}$ and loss $\gamma^{\left(d\right)}$ couple different Keldysh contours in Eq.~\eqref{eq:Lagrangian}, a signature of the non-equilibrium nature of the system.

From now on we focus on the steady state of quadratic lattice models, but transient dynamics and interacting systems can also be studied using this formalism~\cite{Sieberer_2016,Keldysh-manybody1,Keldysh-manybody2,NonlinearPhotonics}. Since the system is time-translation invariant, it is useful to Fourier transform the action to frequency domain and write $H_{\pm} = \sum_{jl} H_{jl} \bar{\psi}_{j,\pm}  \psi_{l,\pm}$. We show below that this frequency-dependence translates to the observables, where $\omega$ physically corresponds to the energy at which the steady state of the system is being probed.

For practical calculations it is useful to perform a Keldysh rotation. In the bosonic case it corresponds to $\psi_{\pm}=\left(\psi_{c}\pm\psi_{q}\right)/\sqrt{2}$, and the bosonic action becomes:
\begin{equation}
S_{b}=\int_{\omega}\Psi^{\dagger}\left(\begin{array}{c|c}
0 & \omega-\mathcal{H}_{A}\\
\hline \omega-\mathcal{H}_{R} & i\Gamma
\end{array}\right)\Psi ,
\label{eq:Keldysh-matrix}
\end{equation}
where we have defined $\int_{\omega}=\int\frac{d\omega}{2\pi}$ and written the fields in vector form $\Psi=(\vec{\psi}_c,\vec{\psi}_q)$, being $\vec{\psi}_{\alpha}=(\psi_{1,\alpha},\psi_{2,\alpha},\ldots)$ and $\alpha=c,q$. The different blocks in Eq.~\eqref{eq:Keldysh-matrix} are given by: $\mathcal{H}_{A/R} = H \pm i \frac{\gamma^{\left(d\right)}-\gamma^{\left(p\right)}}{2}$
and $\Gamma = \gamma^{\left(d\right)}+\gamma^{\left(p\right)}$.
A key observation is that the non-Hermitian matrix $\mathcal{H}_{R}$ correspond to the effective Hamiltonian proposed to study the short-time dynamics of dissipative systems~\cite{ExceptionalPoints1}.

In the fermionic case, the Keldysh rotation is slightly different~\cite{Keldysh-fermions}, but importantly, it changes the sign of the gain contribution in the action. 
This results in the following expression for fermions:
\begin{equation}
S_{f}=\int_{\omega}\bar{\Psi}^{T}
\left(\begin{array}{c|c}
\omega-\mathcal{H}_{R} & i\Gamma\\
\hline 0 & \omega-\mathcal{H}_{A}
\end{array}\right)\Psi\label{eq:FKeldysh-matrix}
\end{equation}
with blocks now given by 
$\mathcal{H}_{A/R} = H \pm i\frac{\gamma^{\left(d\right)}+\gamma^{\left(p\right)}}{2}$
and $\Gamma = \gamma^{\left(d\right)}-\gamma^{\left(p\right)}$.
We show below that the sign change in the pump term will have important consequences in the resulting topological phase diagram.

Finally, to turn the formalism into an effective calculation tool we define the generating functional, from which we can obtain correlation functions by functional differentiation,
\begin{eqnarray}
Z\left[J_{c},J_{q},\bar{J}_{c},\bar{J}_{q}\right] &=& \prod_{l=1}^N \int\mathcal{D}\psi_{l,c} \mathcal{D}\bar{\psi}_{l,c}\mathcal{D}\psi_{l,q} \mathcal{D}\bar{\psi}_{l,q} e^{i S}\\
&&\times e^{i \int_{\omega} \left(\bar{j}_{l,c} \psi_{l,q} + \bar{j}_{l,q}\psi_{l,c}+j_{l,c} \bar{\psi}_{l,q} + j_{l,q}\bar{\psi}_{l,c}\right)}\nonumber
\end{eqnarray}
where we have defined the sources $J_{\alpha} = (j_{1,\alpha},j_{2,\alpha},\ldots)$. 
The final form of the generating functional is obtained by Gaussian integration:
\begin{equation}
Z\left[J,\bar{J}\right]=e^{-i\int_{\omega}\bar{J}^T\left(\omega\right)G\left(\omega\right)J\left(\omega\right)}.
\label{eq:Generating0}
\end{equation}
It is a quadratic form of the sources $J=(J_c, J_q)$, with $G$ obtained from the inverse of the action~\cite{LU2002119} (Eq.~\eqref{eq:Keldysh-matrix} for the bosonic and Eq.~\eqref{eq:FKeldysh-matrix}  for the fermionic case). In general, $G(\omega)$ is a $2 \times 2$ block matrix with entries:
\begin{equation}
G_{A/R}=\frac{1}{\omega-\mathcal{H}_{A/R}},\ G_{K}=G_{R}^{-1}\left(-i\Gamma\right)G_{A}^{-1}\label{eq:Green-function},
\end{equation}
being $G_{A/R}$ the advanced/retarded and $G_K$ the Keldysh Green function. From Eq.~\eqref{eq:Generating0} it is possible to obtain all correlation functions by functional differentiation.

Concretely, here we are interested in 2-point correlation functions of the form
$\mathcal{M}_{jl}(\omega) = \int d\tau \langle \psi_j^\dagger(t) \psi_l(t+\tau) \rangle e^{-i \omega \tau}$, which can be expressed in terms of  Green's functions \cite{Sieberer_2016} as
$\mathcal{M}(\omega) = \frac{i\eta}{2} \left[G^{K}(\omega)+G^{A}(\omega)-G^{R}(\omega)\right]$, where $\eta=\pm 1$ for bosons/fermions. Remarkably, this expression can be simplified in the case of gain/loss systems [see Supplementary Material (SM)]:
\begin{equation}
\mathcal{M}(\omega)   = G_R(\omega) \gamma^{(p)} G_A(\omega).
    \label{eq:corr}
\end{equation}
Notice that the last term in Eq.~\eqref{eq:corr} is independent of the particle statistics and relates the two-point correlations with the non-Hermitian matrices $\mathcal{H}_{A/R}$ and with the incoherent pump of particles in the system $\gamma^{(p)}$.

{\it Topological properties.---} 
We address now the topological characterization in terms of $G_R(\omega)$, 
which has also been considered as a topological tool in different situations~\cite{PhysRevB.83.085426,PhysRevB.86.165116,PhysRevX.2.031008,PhysRevLett.124.056802} and is related to the electromagnetic response in topological field theories~\cite{PhysRevLett.126.216405,PhysRevLett.124.240404,Wang629}. 

For that, we first define the doubled Hamiltonian $\tilde{\mathcal{H}}(\omega)$:
\begin{equation}
\tilde{\mathcal{H}}(\omega)=
\left(\begin{array}{cc}
0 & \omega-\mathcal{H}_{R}\\
\omega-\mathcal{H}_{A} & 0
\end{array}\right),
\label{eq:doubled-H}
\end{equation}
which is Hermitian by construction (notice that $\mathcal{H}_R^\dagger = \mathcal{H}_A$) and has a built-in chiral symmetry due to its block structure.  
The doubled Hamiltonian has been used as a formal technique in the classification of topological phases of non-Hermitian systems with the poing-gap criterion 
\cite{PhysRevX.8.031079,PhysRevLett.122.143901,PhysRevX.9.041015,Periodic-table-Non-hermitian} 

In this work, $\tilde{\mathcal{H}}$ will allow us to link Hermitian topological invariants and the non-equilibrium Green's functions.
This is because $\tilde{\mathcal{H}}$ can serve us to compute the inverse of $\omega-\mathcal{H}_R$ \cite{PhysRevA.103.033513}, with the advantage that its eigenvalues are insensitive to the skin effect~\cite{PhysRevLett.124.086801,Skin-effect1,Quantum-Anomaly}. To see this, notice that due to the artificial chiral symmetry, the eigenstates of 
$\tilde{\mathcal{H}}$ can be written as 
$\tilde{\mathcal{H}} 
\left(
\begin{array}{c}
u_n \\
\pm v_n 
\end{array}\right)
= \pm \tilde{\epsilon}_n
\left(
\begin{array}{c}
u_n \\
\pm v_n 
\end{array}\right)
$, with $\tilde{\epsilon}_n > 0$. 
For example, 
in 1D, if $\tilde{\cal H}$ is in a topologically non-trivial phase, zero-energy states, will have $\tilde{\epsilon}_n \approx 0$, and the corresponding vectors $u_n$, $v_n$ are left and right localized edge-states. 
By using this observation we can easily derive (see SM):
\begin{equation}
G_R(\omega)_{jl} = \sum_n \frac{1}{\tilde{\epsilon}_n} (v_n)_j (u_n)_l^* ,
\label{eq:GR_uv}
\end{equation}
which relates the non-Hermitian Green's function 
and the eigenstates of the doubled Hamiltonian. 
Importantly, Eq.~\eqref{eq:GR_uv} indicates that topological zero-energy modes of the doubled Hamiltonian, $\tilde{{\cal H}}(\omega)$, dominate the correlation functions [c.f. Eq.~\eqref{eq:corr}], through the inverse energy factor, $1/\tilde{\epsilon}_n$. 
This relation between $\tilde{\mathcal{H}}$ and $G_R(\omega)$ shows that one can perform the topological analysis of $\tilde{\mathcal{H}}$ in terms of the tenfold way~\cite{Ryu_2010} applied to chiral symmetric Hamiltonians, and directly link its topological phases with physical observables determined by $G_R(\omega)$. 
This classification results in a smaller number of different phases than those predicted in~\cite{PhysRevX.9.041015}, since it is restricted to those topological properties that are directly related to observables of the quantum open lattice.

\begin{figure}
    \centering
    \includegraphics[width = 0.9\linewidth]{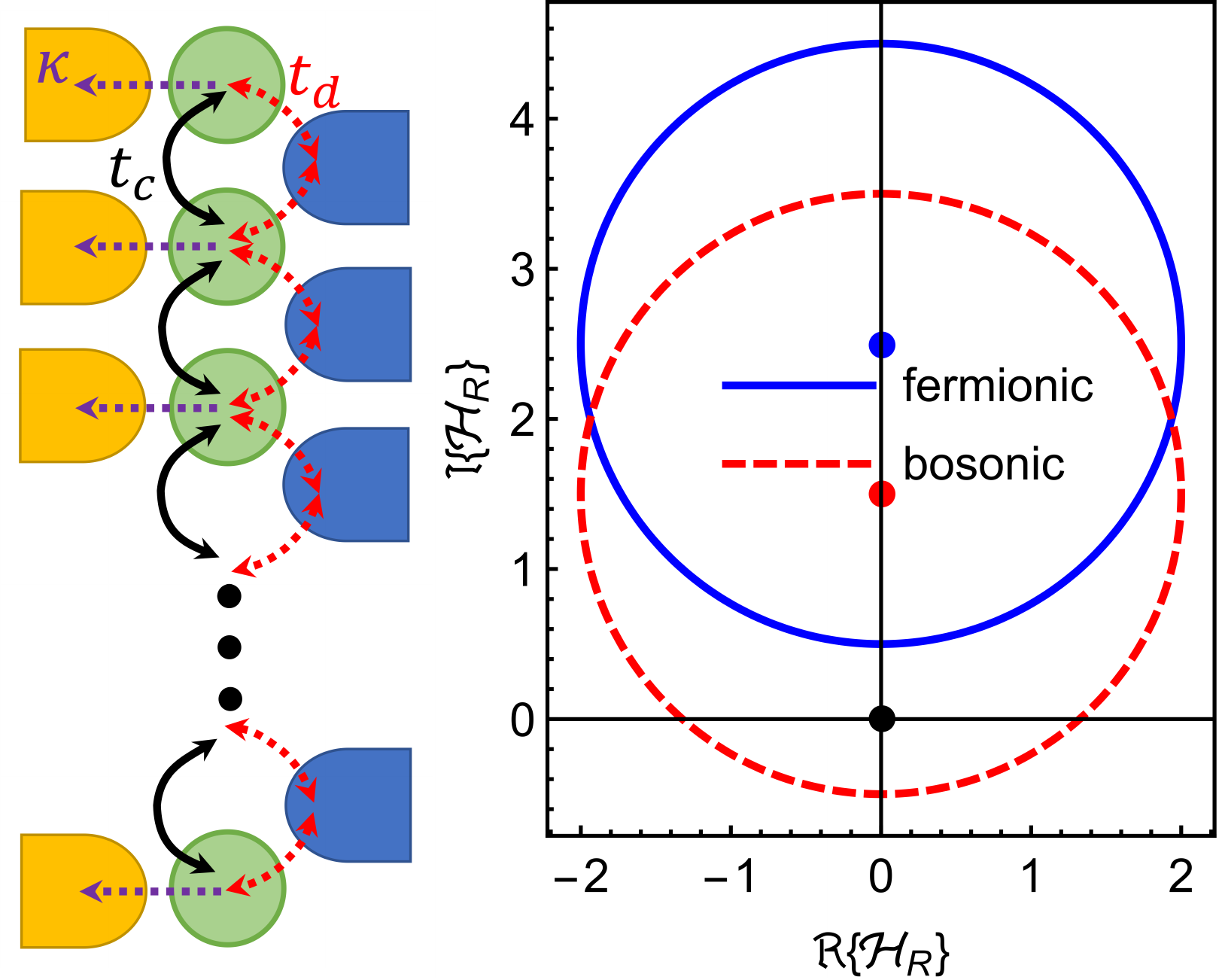}
    \caption{(Left)  Schematic for the H-N model. Sites in the array coherently couple with hopping $t_c$ and auxiliary sites are dissipatively coupled via $\kappa$ and $t_d$. (Right) Complex plane plot of the eigenvalues of $\mathcal{H}_R$ for the bosonic(red) and the fermionic(blue) H-N model with PBC. The fermionic case never encloses the origin and remains trivial. Red and blue dots show the collapse of the eigenvalues due to the skin effect for OBC.}
    \label{fig:Figure1}
\end{figure}

{\it Hatano-Nelson model.---} 
The Hatano-Nelson (H-N) model is a canonical example of topology induced by dissipation~\cite{PhysRevLett.77.570,PhysRevX.8.031079,Longhi15}. 
In the bosonic version, the Hamiltonian $\hat{H}=\sum_{i,j}t_{i,j}\hat{a}_{i}^{\dagger}\hat{a}_{j}$ describes hopping in a lattice with a background gauge field $\phi$, where $t_{i,j}=\omega_{0}\delta_{i,j}+t_{c}\left(e^{i\phi}\delta_{i,j-1}+e^{-i\phi}\delta_{i,j+1}\right)$ and $\omega_{0}$ describes the detuning from the cavity frequency. 
In addition, the particle dynamics is influenced by local loss $\gamma_{i,j}^{\left(d\right)}=\kappa\delta_{i,j}$ and non-local gain $\gamma_{i,j}^{\left(p\right)}=4t_{d}\delta_{i,j}+2t_{d}\left(\delta_{i,j-1}+\delta_{i,j+1}\right)$ [see schematic in Fig.\ref{fig:Figure1}(left)].
The implementation of the bosonic model can be carried out, for example, by using reservoir engineering and Floquet techniques for inducing synthetic gauge fields \cite{Metelmann15,Metelmann17,PhysRevLett.122.143901}.

From Eq.~\eqref{eq:Keldysh-matrix} it is straightforward to write the different blocks of the bosonic action $S_b$, for the case of periodic and open boundary conditions (PBC and OBC, respectively). In the case of PBC each block corresponds to:
\begin{align}
\mathcal{H}_{A/R}= & \omega_{0}+2t_{c}\cos\left(k-\phi\right) \pm i\frac{\kappa-8t_{d}\cos^{2}\left(\frac{k}{2}\right)}{2},
\label{eq:HAR}
\\
\Gamma= & \kappa+8t_{d}\cos^{2}\left(\frac{k}{2}\right) .
\end{align}
According to the standard classification of non-hermitian matrices, $\mathcal{H}_R$ belongs to the AI class because it lacks all symmetries~\cite{PhysRevX.9.041015}. In consequence, its winding number is non-zero when the complex eigenvalues form a point-gap which encloses the origin [see Fig.~\ref{fig:Figure1} (right)]. 
Importantly, the classification of $\mathcal{H}_R$ is $\omega$-independent and only indicates the presence of a topological amplification phase, neglecting the range of $\omega$ where states are amplified.

If we instead classify the doubled Hamiltonian $\tilde{\mathcal{H}}$ using the 10-fold way, we find that it belongs to the AIII class due to the artificial chiral symmetry. Its topological phase is characterized by a winding number which can be written as:
\begin{equation}
W_{1}\left(\omega\right)=\int_{-\pi}^{\pi}\frac{dk}{2\pi i}\partial_{k}\log\left(\omega-\mathcal{H}_{R}\right),
\label{eq:Winding1}
\end{equation}
Notice that its $\omega$-dependence naturally arises and is physically motivated by the fact that $\tilde{\mathcal{H}}$ can be used to compute the inverse of $\mathcal{H}_R-\omega$, which controls the behavior of the two-point functions. 
In addition, this is in agreement with the $\omega$-dependent topological invariants predicted in dissipative systems~\cite{Top-Winding-finite,PhysRevLett.126.216405,PhysRevA.103.033513}.
\begin{figure}
    \centering
    \includegraphics[width=0.9\linewidth]{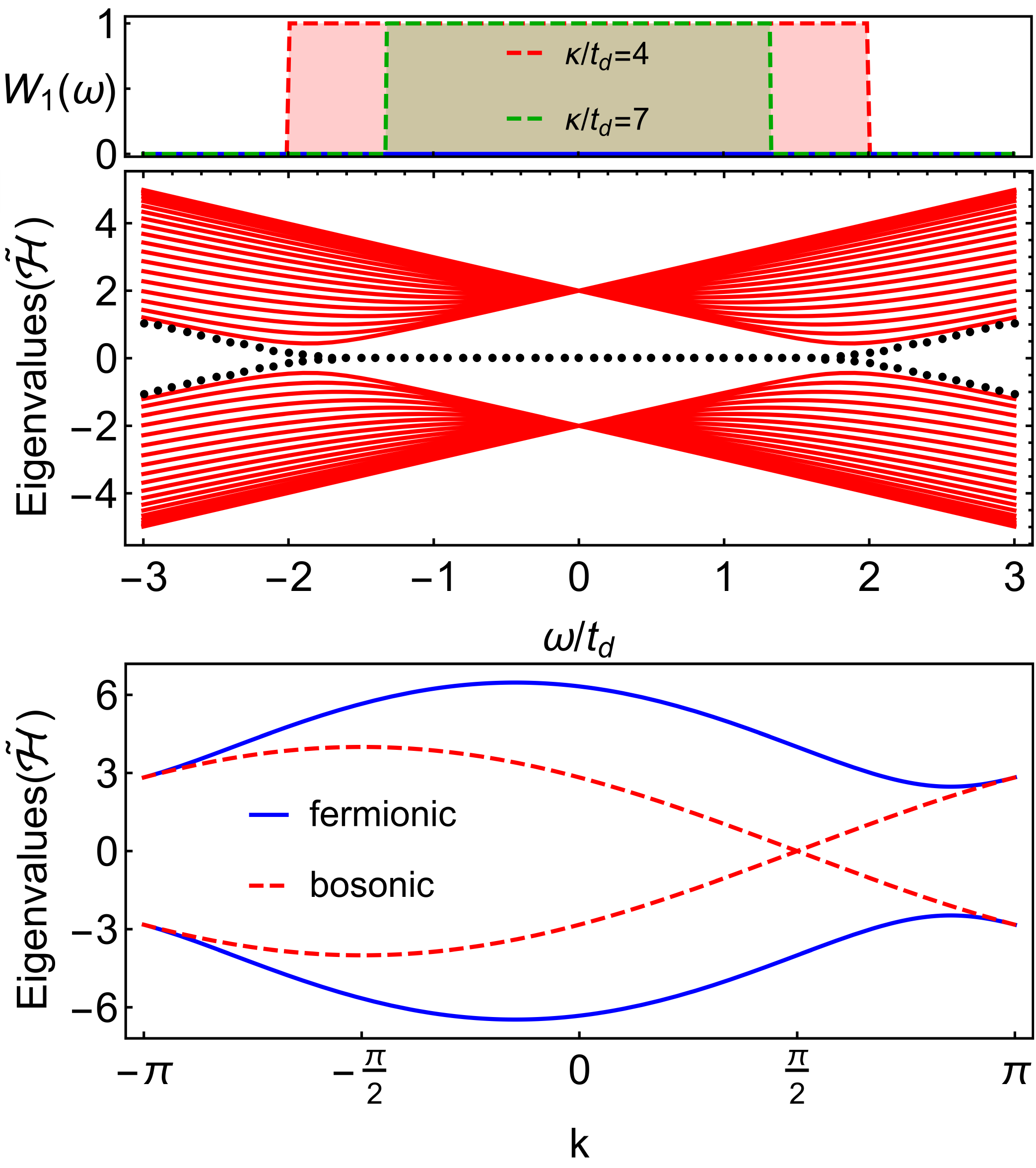}
    \caption{(Top) $W_1(\omega)$ for different values of $\kappa/t_d$. In the fermionic case $W_1(\omega)$ is always zero (blue). (Middle) Eigenvalues of $\tilde{\mathcal{H}}$ vs $\omega$ for $\kappa/t_d$=4. The spectrum for PBC is shown in red, while black dots indicate the two boundary modes with OBC. (Bottom) Eigenvalues of $\tilde{\mathcal{H}}$ vs $k$ for $\kappa/t_d$=4 and $\omega=2$. All plots consider $t_c/t_d=1$ and $\phi=\pi/2$}
    \label{fig:Figure2}
\end{figure}
Fig.~\ref{fig:Figure2}(top) shows the value of $W_1(\omega)$ for different loss rates $\kappa$, which affects the range of frequencies which can be amplified.
Fig.~\ref{fig:Figure2}(middle) shows the eigenvalues of $\tilde{\mathcal{H}}$ for PBC (red) with the appearance of a pair of topological boundary modes for OBC (black dots). The lack of skin effect in the eigenvalues of  $\tilde{\mathcal{H}}$ and the match between the appearance of boundary modes and changes in the $W_1(\omega)$ are obvious advantages with respect to $\mathcal{H}_R$.

In analogy with Hermitian topology we can also see in Fig.~\ref{fig:Figure2} (bottom) that for PBC, the eigenvalues of $\tilde{\mathcal{H}}$ vs $k$ show that the critical point is linked with a gap closure in the bosonic case.

Physically, the topological phase in the bosonic H-N model corresponds to unidirectional amplification. The $\omega$-dependence in $W_1(\omega)$ is crucial, as it indicates that topological amplification happens for a finite range of frequencies only. This is interesting to relate topology in dissipative systems with its experimental detection. The simplest way consists in detecting the number of particles at each site $\langle a_j^\dagger a_j\rangle$, which in the amplification phase shows an exponential dependence with the array length~\cite{PhysRevA.103.033513}. This however does not characterize the $\omega$-dependence of $W_1(\omega)$, even if the number of particles is measured at different $\omega$, because there is not a sharp transition as a function of $\omega$ (the gap closes continuously). 

An alternative approach, for example, is to measure the response function to a perturbation of the frequency at site $l$, by adding the term 
$H_{\rm I} = \Omega_l \hat{\psi}^\dagger_l \hat{\psi}_l$ to the Hamiltonian. 
We define a susceptibility related to the variation of the excitation number at frequency $\omega$ at another site $j$, 
$    \chi_{jl}(\omega) = 
    d \langle n_j(\omega) \rangle / d \Omega_l $, and find 
  (see SM):
\begin{equation}
    \chi_{jl}(\omega) = G_{jl}^R(\omega) \mathcal{M}_{lj}(\omega) + \mathcal{M}_{jl}(\omega) G_{lj}^A(\omega) .
\end{equation}
\begin{figure}
    \centering
    \includegraphics[width=0.9\linewidth]{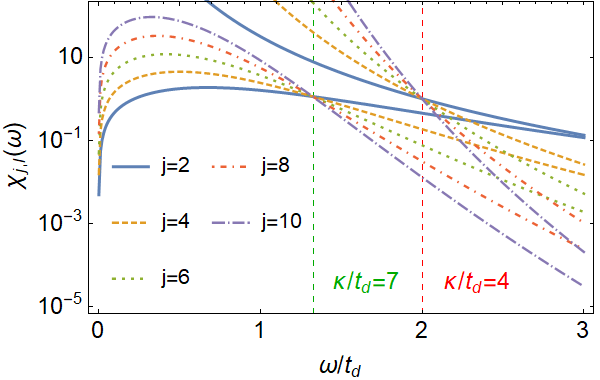}
    \caption{Logarithmic plot of $\chi_{j,l}(\omega)$ between sites $l=1$ and $j=2,4,6,8\text{ and }10$. We have considered the bosonic case for an array with $N=10$ sites, $t_c/t_d=1$ and $\phi=\pi/2$. The crossing at a certain value of $\omega$ allows to extract the position of the critical point, indicated for the cases $\kappa/t_d=4\text{ and }7$.}
    \label{fig:Figure3}
\end{figure}
Interestingly, Fig.~\ref{fig:Figure3} shows that plotting in logarithmic scale the susceptibility between different sites allows to directly detect the critical point. This is a consequence of the topological phase transition to unidirectional amplification, where signals are exponentially amplified with the number of sites and  their scaling when measured at different sites is drastically affected. Importantly, this indirect detection gives very accurate results, even for small arrays (note Fig.~\ref{fig:Figure3} is calculated for a system with only $N=10$ sites).

{\it Fermions vs bosons.---} The symmetry class of $\mathcal{H}_R$ does not depend on whether we consider a fermionic or a bosonic lattice, however, we can show that particle statistics drastically affects topological phases. 
The key observation is the sign change in the pump term in 
the fermionic case, Eq.~\eqref{eq:FKeldysh-matrix}, which physically accounts for Pauli exclusion as opposed to bosonic amplification. 
The consequences of this for the H-N model can be derived from Eq.~\eqref{eq:HAR}, which in the fermionic case leads to ${\rm Im} \left( H_R \right) 
\propto - \kappa - 8 t_d \cos^2(k/2)$. 
Since $\kappa, \ t_d > 0$, 
${\rm Im} \left( H_R \right)$ does not change sign, which is a necessary condition for $W_1 (\omega) \neq 0$ 
[see Fig.~\ref{fig:Figure1}(right)]. Its consequences are also shown in Fig.~\ref{fig:Figure2}, where the winding number is always zero and the band structure remains always gaped. We thus conclude that the fermionic H-N model has a topologically trivial phase diagram.

Formally, the limitations found in the fermionic H-N model could be surpassed if the diagonal and non-diagonal elements of the matrix $\gamma^{(p)}$ could be independently tuned, which would free the model from the $\cos^2(k/2)$ dependence in ${\cal H}_{R}$. However, from the derivation of the H-N master equation it can be shown that $\gamma^{(p)}_{j j} = 2 \gamma^{(p)}_{j,j+1}$ (see SM), which accounts for the fact that dissipative couplings between sites induced by a common bath inevitably come together with local dissipation terms. This result has a clear physical meaning, since the directional amplification that would result from a non-trivial topological phase is not expected to occur in a fermionic lattice.

{\it Conclusions and Outlook.---} 
We have used the Keldysh formalism to connect topological properties of non-Hermitian matrices with physical observables of quantum open systems. 
In particular, we have defined a frequency dependent topological index that can be used to characterize properties of non-equilibrium Green's functions.
Our formalism allows us to obtain a unified description of bosonic and fermionic open models, and we have unveiled fundamental differences between the topological phases of the two cases. 
We have applied our theoretical framework to the 1D Hatano-Nelson model, and we have explicitly shown how physical observables and response functions of that model can be used to detect non-trivial topological phases. Our theory leads to an unambiguous definition of topological phases and topological phase transitions in quantum open systems. 

Our work paves the way for further applications of the Keldysh theoretical machinery~\cite{Sieberer_2016} in the description of topological gain/loss systems. In particular, adding interactions to the theoretical framework presented here would allow us to investigate topological interacting quantum open systems. One can also apply our ideas to transient physics rather than to the steady-state ~\cite{Viola1,Viola3}. From a practical point of view, our work can lead to the design of quantum metrology or sensing protocols~\cite{Sensing1,Sensing-McDonald2020,Sensing-Koch2021}, by exploiting the extreme sensitivity of the system to input fields and perturbations in non-trivial topological phases. Our results are relevant for current experimental setups in photonic lattices where the H-N model could be implemented using Floquet techniques and reservoir engineering. Similar techniques may lead to the investigation of fermionic models by using, for example, arrays of coupled quantum dots~\cite{Quantumdots1,Quantumdots2}.

\begin{acknowledgments}
We acknowledge financial support from the Proyecto Sinérgico CAM 2020 Y2020/TCS-6545 (NanoQuCo-CM), the CSIC Research Platform on Quantum Technologies PTI-001 and from Spanish project PGC2018-094792-B-100(MCIU/AEI/FEDER, EU). T.R. further acknowledges support from the EU Horizon 2020 program under the Marie Sk\l{}odowska-Curie grant agreement No. 798397, and from the Juan de la Cierva fellowship IJC2019-040260-I.
\end{acknowledgments}

\begin{widetext}

\appendix

\section{Observables:}
Here we detail the derivation of the expression for the general 2-point function in the main text, for the cases of both, bosons and fermions dissipative systems.
Let us begin with the general definition of the advanced, retarded and Keldysh Green functions, in terms of the lesser, greater, time-ordered and anti time-ordered ones:
\begin{align}
    G^{R}\left(t,t^{\prime}\right)=& \theta\left(t-t^{\prime}\right)\left(G^{>}\left(t,t^{\prime}\right)-G^{<}\left(t,t^{\prime}\right)\right)\\
G^{A}\left(t,t^{\prime}\right)=& \theta\left(t^{\prime}-t\right)\left(G^{<}\left(t,t^{\prime}\right)-G^{>}\left(t,t^{\prime}\right)\right)\\
G^{K}\left(t,t^{\prime}\right)=& G^{>}\left(t,t^{\prime}\right)+G^{<}\left(t,t^{\prime}\right)
\end{align}

These equalities are valid in bosonic and fermionic systems. In addition, the definition of the Green functions in the Keldysh contour is also valid for both instances:
\begin{align}
    G^{<}\left(t,t^{\prime}\right)=& -i\langle\psi_{+}\left(t\right)\bar{\psi}_{-}\left(t^{\prime}\right)\rangle\\
G^{>}\left(t,t^{\prime}\right)=& -i\langle\psi_{-}\left(t\right)\bar{\psi}_{+}\left(t^{\prime}\right)\rangle\\
G^{T}\left(t,t^{\prime}\right)=& -i\langle\psi_{+}\left(t\right)\bar{\psi}_{+}\left(t^{\prime}\right)\rangle\\
G^{\bar{T}}\left(t,t^{\prime}\right)=& -i\langle\psi_{-}\left(t\right)\bar{\psi}_{-}\left(t^{\prime}\right)\rangle
\end{align}

The differences between particles become evident when we connect these Green functions with their respective many-body expressions:
\begin{align}
    G^{<}\left(t,t^{\prime}\right)=& -\eta i\langle\hat{\psi}^{\dagger}\left(t^{\prime}\right)\hat{\psi}\left(t\right)\rangle\\
    G^{>}\left(t,t^{\prime}\right)=& -i\langle\hat{\psi}\left(t\right)\hat{\psi}^{\dagger}\left(t^{\prime}\right)\rangle\\
G^{T}\left(t,t^{\prime}\right)=& -i\theta\left(t-t^{\prime}\right)\langle\hat{\psi}\left(t\right)\hat{\psi}^{\dagger}\left(t^{\prime}\right)\rangle-i\eta\theta\left(t^{\prime}-t\right)\langle\hat{\psi}^{\dagger}\left(t^{\prime}\right)\hat{\psi}\left(t\right)\rangle\\
G^{\bar{T}}\left(t,t^{\prime}\right)=& -i\eta\theta\left(t-t^{\prime}\right)\langle\hat{\psi}^{\dagger}\left(t^{\prime}\right)\hat{\psi}\left(t\right)\rangle-i\theta\left(t^{\prime}-t\right)\langle\hat{\psi}\left(t\right)\hat{\psi}^{\dagger}\left(t^{\prime}\right)\rangle
\end{align}

where $\eta=\pm 1$ for bosons/fermions, respectively. The sign difference due to $\eta$ leads to the commutator/anti-commutator in the advanced and retarded Green's function, for the bosonic/fermionic case:
\begin{align}
    G^{R}\left(t,t^{\prime}\right)=& -i\theta\left(t-t^{\prime}\right)\langle\left[\hat{\psi}\left(t\right),\hat{\psi}^{\dagger}\left(t^{\prime}\right)\right]_{-\eta}\rangle\\
G^{A}\left(t,t^{\prime}\right)=& i\theta\left(t^{\prime}-t\right)\langle\left[\hat{\psi}\left(t\right),\hat{\psi}^{\dagger}\left(t^{\prime}\right)\right]_{-\eta}\rangle\\
G^{K}\left(t,t^{\prime}\right)=& -i\langle\left[\hat{\psi}\left(t\right),\hat{\psi}^{\dagger}\left(t^{\prime}\right)\right]_{\eta}\rangle
\end{align}
being $[...,...]_{-}$ the ordinary commutator and $[...,...]_{+}$ the anti-commutator.

Now, we are interested in 2-point functions of the form $\langle\hat{\psi}^{\dagger}\left(t+\tau\right)\hat{\psi}\left(t\right)\rangle$, which in frequency space can be obtained in terms of the Green functions from their many-body representation:

\begin{align}
    G^{R}\left(t,t+\tau\right)=& i\theta\left(-\tau\right)\left(\eta\langle\hat{\psi}^{\dagger}\left(t+\tau\right)\hat{\psi}\left(t\right)\rangle-\langle\hat{\psi}\left(t\right)\hat{\psi}^{\dagger}\left(t+\tau\right)\rangle\right)\\
G^{A}\left(t,t+\tau\right)=& -i\theta\left(\tau\right)\left(\eta\langle\hat{\psi}^{\dagger}\left(t+\tau\right)\hat{\psi}\left(t\right)\rangle-\langle\hat{\psi}\left(t\right)\hat{\psi}^{\dagger}\left(t+\tau\right)\rangle\right)\\
G^{K}\left(t,t+\tau\right)=& -\eta i\langle\hat{\psi}^{\dagger}\left(t+\tau\right)\hat{\psi}\left(t\right)\rangle-i\langle\hat{\psi}\left(t\right)\hat{\psi}^{\dagger}\left(t+\tau\right)\rangle
\end{align}

Some manipulations lead to the following expression, which can be checked by inserting the previous definitions:
\begin{equation}
    \langle\hat{\psi}^{\dagger}\left(t+\tau\right)\hat{\psi}\left(t\right)\rangle=	\eta\frac{i}{2}\left[G^{K}\left(t,t+\tau\right)+G^{A}\left(t,t+\tau\right)-G^{R}\left(t,t+\tau\right)\right]
\end{equation}

If we now Fourier transform and insert the definitions of $\mathcal{H}_{A/R}$ and $\Gamma$ for the bosonic or the fermionic actions, we find:
\begin{align}
    \langle\hat{\psi}^{\dagger}\hat{\psi}\rangle \left(t,\omega\right)=& \eta\frac{i}{2}\int_{-\infty}^{\infty}d\tau e^{-i\omega\tau}\left\{ G^{K}\left(t,t+\tau\right)+G^{A}\left(t,t+\tau\right)-G^{R}\left(t,t+\tau\right)\right\}\nonumber\\
=& \eta\frac{i}{2}\left[G^{K}\left(\omega\right)+G^{A}\left(\omega\right)-G^{R}\left(\omega\right)\right]\nonumber\\
=&\eta\frac{i}{2}\frac{1}{\omega-\hat{\mathcal{H}}_{R}}\left(\hat{\mathcal{H}}_{A}-\hat{\mathcal{H}}_{R}-i\hat{\Gamma}\right)\frac{1}{\omega-\hat{\mathcal{H}}_{A}}\nonumber\\
=& \frac{1}{\omega-\hat{\mathcal{H}}_{R}}\hat{\gamma}_{p}\frac{1}{\omega-\hat{\mathcal{H}}_{A}}
\end{align}
which indicates that the 2-point function expressed in terms of the matrix elements of the action is the same for both types of particles.

\section{Relating $G_R(\omega)$ and $\tilde{{\cal H}}(\omega)$}
In this section we prove Eq. \eqref{eq:GR_uv}. 
Let us recall that $G_R(\omega) = 1/(\omega - {\cal H}_R)$. 
Due to the chiral symmetry of $\tilde{\cal H}(\omega)$, we can write its eigenvalues in the form,
\begin{equation}
\tilde{\mathcal{H}} 
\left(
\begin{array}{c}
u_n \\
\pm v_n 
\end{array}\right)
= \pm \tilde{\epsilon}_n
\left(
\begin{array}{c}
u_n \\
\pm v_n 
\end{array}\right),
\label{eq:eigen}
\end{equation}
where $u_n$, $v_n$ are $n = 1, \dots, N$ vectors of dimension $N$ (with $N$ the number of sites of the lattice), forming two orthonormal basis.
 
Let us define the unitary matrices,
\begin{equation}
    {U}_{nj} = (u_n)_j, \ \ V_{nj} = (v_n)_j ,
\end{equation}
as well as the diagonal matrix
\begin{equation}
    {S}_{n m} = \tilde{\epsilon}_n \delta_{n m}. 
\end{equation}
Matrices $U$, $V$, $S$ can alternatively be obtained from the singular value decomposition of $\omega - {\cal H}_R$. 
Eq. \eqref{eq:eigen} can be rewritten in matrix form,
\begin{equation}
\tilde{\mathcal{H}}(\omega)=
\left(\begin{array}{cc}
0 & U S V^\dagger \\
V S U^\dagger & 0
\end{array}\right) .
\label{eq:doubled-H-svd}
\end{equation}
This leads to the identities (via Eq. \eqref{eq:doubled-H}),
\begin{eqnarray}
    \omega - {\cal H}_R  &=& U S V^\dagger \nonumber \\
    G_R(\omega) &=& (\omega - {\cal H}_R)^{-1}  = V^\dagger S^{-1} U,
\end{eqnarray}
which finally allow us to write Eq. \eqref{eq:GR_uv} of the main text.

\section{Calculation of the winding number}

The calculation of the winding number for the N-H model can be done
analytically using contour techniques. For that we consider the integral:
\begin{align*}
W_{1} & =\int_{-\pi}^{\pi}\frac{dk}{2\pi i}\partial_{k}\log\left(\omega-\mathcal{H}_{R}\right)\\
 & =i\oint\frac{dz}{2\pi}\frac{t_{-}t_{+}^{-1}-z^{2}}{z\left(z-z_{+}\right)\left(z-z_{-}\right)}
\end{align*}
and notice that it has three poles at coordinates:
\[
z_{0}=0,\ iz_{\pm}=\frac{\alpha}{2t_{+}}\pm\frac{1}{2t_{+}}\sqrt{\alpha^{2}+4t_{+}t_{-}}
\]
where we have defined $z=e^{ik}$, $t_{\pm}=t_{c}e^{\pm i\phi}+it_{d}$
and $\alpha=2t_{d}-\kappa/2-i\left(\omega_{0}-\omega\right)$. For
the topological phase all the poles are inside the unit circle, and
for the trivial only $z_{+}$ and $z_{0}$ remain inside. As the contour
integral contributes as $-1$ for the $z_{0}$ pole, and $+1$ for
the $z_{\pm}$ poles, we find that the winding number is given by
the condition for the modulus of $z_{+}$ to be smaller than one.
Hence we can write the winding number as:
\begin{equation}
W_{1}=\theta\left(1-\left|z_{+}\right|\right)
\end{equation}

\section{Susceptiblity}
We are interested in response functions. They characterize changes in local observables, when small variations of the parameters are produced at a different locations of the system. In analogy with magnetic systems, where the magnetic susceptibility detects changes in the magnetization as the external field is tuned, here we study variations in the number of particles as we slightly modify one of the local frequencies. This can be formally addressed by adding a local frequency shift to the Hamiltonian:
\begin{equation}
\hat{H} \to \hat{H} + \Omega_l \hat{\psi}^\dagger_l \hat{\psi}_l,
\end{equation}
and calculating the following response function, which can be understood as a particle number susceptibility:
\begin{equation}
    \chi_{jl}(\omega) = 
    \frac{d \langle n_j(\omega) \rangle}{d \Omega_l} 
\end{equation}
where $\langle n_j(\omega) \rangle$ is the average number of particles at frequency $\omega$. Note that this is very relevant for quantum metrology/sensing applications, since measuring variations in local frequency shifts is the working principle of many devices. 

The calculation is relatively easy in the perturbative limit, we only have to keep in mind the following differentiation rule for matrices:
\begin{equation}
\frac{d}{d h} \left( X - h \Delta  \right)^{-1} = 
 X^{-1} \Delta X^{-1}.
\end{equation}
Concretely in our case, we need to calculate the following:
\begin{eqnarray}
    \chi_{jl}(\omega)&=&\frac{d}{d\Omega_l}\langle \hat{\psi}_j^{\dagger} \hat{\psi}_j \rangle (\omega) = \left( G_R (\omega) \mathbf{1}_l \langle \hat{\psi}^\dagger \hat{\psi} \rangle (\omega)+ \langle \hat{\psi}^\dagger \hat{\psi} \rangle (\omega) \mathbf{1}_l G_A (\omega) \right)_{jj}\\
    &=& 2\text{Re}\left[G_{jl}^R(\omega)\langle \hat{\psi}_l^\dagger \hat{\psi}_j \rangle(\omega)\right]
\end{eqnarray}
where $\langle \hat{\psi}^\dagger \hat{\psi} \rangle(\omega)$ is the matrix of all two-point functions, we have used $G_A(\omega)=G_R^{\dagger}(\omega)$ and that the derivative of the Green function matrix is
\begin{equation}
\frac{\partial}{\partial \Omega_l} G_R(\omega) = 
\frac{\partial}{\partial \Omega_l} 
\frac{1}{\omega -  \Omega_{l} {\bf 1}_{l} - {\cal H}_R } |_{\Omega_l = 0}= 
G_R(\omega) {\bf 1}_{l} G_R(\omega),
\label{eq:derivative_GR}
\end{equation}
where ${\bf 1}_{l}$ is a matrix with all zeros expect at the $l$'th position on the diagonal. This expression is interesting from the point of view of propagation of the perturbation, and it agrees with our physical intuition of the meaning of the retarded Green function.
\begin{figure}
    \centering
    \includegraphics[scale=0.4]{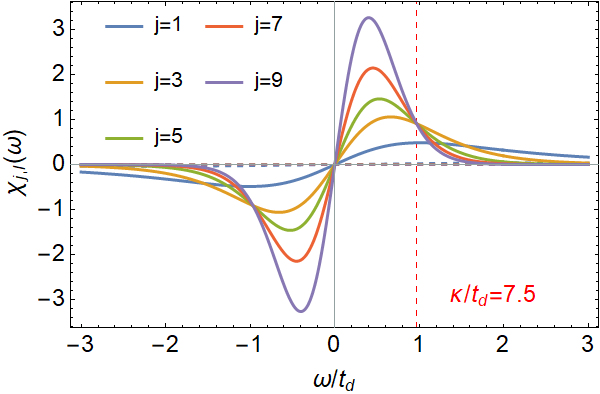}
    \includegraphics[scale=0.4]{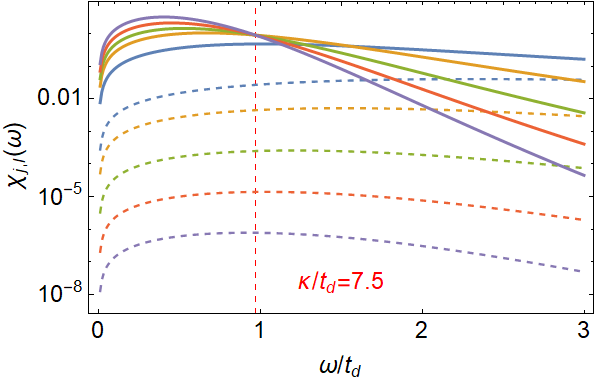}
    \caption{(Left) Susceptibility between sites $l=1$ and $j$ as a function of $\omega$ for the topological ($\kappa/t_d=7.5$) and the trivial ($\kappa/t_d=12$) phase of the Hatano-Nelson model. Dashed lines correspond to the trivial phase while solid correspond to the topological one. Clearly, amplification produces a large scale difference between the two. (Right) Logarithmic plot of the susceptibility, where the $\omega$-dependence of the winding number leads to a crossover region separating the topological and the trivial phases (indicated by a vertical dot-dashed red line). Notice that the crossing (critical point) can be modeled by a set of linear equations: $\log(\chi_{x,y})=\alpha-x(\omega-\beta)$, where $\beta$ fixes the frequency at which the crossing happens and is directly related with the topological critical point. Furthermore, we can separate three contributions in the log and use their dependence on frequency and position to extract their value from experiments: i)$\alpha$ which is independent of both, frequency and position, ii)$\beta$, which is independent of the frequency, and iii)$x \omega$ which is linear in both}
    \label{fig:susceptibility}
\end{figure}
The calculation of the susceptibility shows that due to amplification, there is a large difference between the topological and the trivial phase, which increases as one moves towards the boundary. This is shown in Fig.~\ref{fig:susceptibility}(left), where one can see a difference in orders of magnitude between the two phases. In addition, the off-diagonal susceptibility (i.e., $j\neq l$) is also amplified and Fig.~\ref{fig:susceptibility}(right) shows that a logarithmic plot of the susceptibility can be used to extract the critical point separating the topological and the trivial phase.
\section{Relation between local and non-local dissipative terms}
In order to find the relation between the different dissipative terms, we follow the standard derivation of the quantum master equation, but include the possibility of hopping to a bath reservoir from different sites of the central system:
\begin{equation}
    \hat{V}=\sum_{k\in\text{BZ}}\sum_{n,m}\left[c_{n,m}\left(k\right)\hat{\psi}_{n}^{\dagger}\hat{b}_{k,m}+c_{n,m}^{\ast}\left(k\right)\hat{b}_{k,m}^{\dagger}\hat{\psi}_{n}\right],
\end{equation}
where $\hat{b}_{k,m}$ is the $m$-th bath operator with momentum $k$. This will allow us to model the presence of non-local dissipative hopping in both, bosonic and fermionic models.

The calculation in the interaction picture is more involved in this case, because of our choice for $\hat{V}$, but as we are interested in the coefficients characterizing the dissipative terms rather than in solving the master equation, we can calculate them in the Schrödinger picture and extract them from the integro-differential equation. 

If we consider the usual Born approximation and assume that the density of the environment is in thermal equilibrium, this leads to the integro-differential equation:
\begin{eqnarray}
\partial_{t}\hat{\rho}_{s}\left(t\right)&\simeq&-i\left[\hat{H}_{S},\hat{\rho}_{s}\left(0\right)\right]-\int_{0}^{t}d\tau\left[\hat{H}_{S},\left[\hat{H}_{S},\hat{\rho}_{s}\left(\tau\right)\right]\right]\nonumber\\
&&+\int_{0}^{t}d\tau\sum_{n,n^{\prime}}\left(2\hat{\psi}_{n}^{\dagger}\hat{\rho}_{s}\left(\tau\right)\hat{\psi}_{n^{\prime}}-\left\{ \hat{\psi}_{n^{\prime}}\hat{\psi}_{n}^{\dagger},\hat{\rho}_{s}\left(\tau\right)\right\} \right)G_{n,n^{\prime}}\nonumber\\
&&+\int_{0}^{t}d\tau\sum_{n,n^{\prime}}\left(2\hat{\psi}_{n^{\prime}}\hat{\rho}_{s}\left(\tau\right)\hat{\psi}_{n}^{\dagger}-\left\{ \hat{\psi}_{n}^{\dagger}\hat{\psi}_{n^{\prime}},\hat{\rho}_{s}\left(\tau\right)\right\} \right)P_{n,n^{\prime}},
\end{eqnarray}
where we have ignored the shifts in energy produced by the dissipative terms and defined:
\begin{eqnarray}
G_{n,n^{\prime}}&=&\Re\sum_{k,k^{\prime}}\sum_{m,m^{\prime}}c_{n,m}\left(k\right)c_{n^{\prime},m^{\prime}}^{\ast}\left(k^{\prime}\right)\textrm{Tr}_{B}\left\{ b_{k^{\prime},m^{\prime}}^{\dagger}b_{k,m}\rho_{B}\right\} \\
P_{n,n^{\prime}}&=&\Re\sum_{k,k^{\prime}}\sum_{m,m^{\prime}}c_{n,m}\left(k\right)c_{n^{\prime},m^{\prime}}^{\ast}\left(k^{\prime}\right)\textrm{Tr}_{B}\left\{ b_{k,m}b_{k^{\prime},m^{\prime}}^{\dagger}\rho_{B}\right\} 
\end{eqnarray}
We can now explicitly evaluate the dissipative terms for the case of each bath coupling only to a given pair of sites in the array (nearest neighbors). For this, we fix
$c_{n,m}\left(k\right)=f\left(k\right)\left(\delta_{n,m}+\delta_{n,m+1}\right)$ and evaluate $G_{n,n^{\prime}}$ and $P_{n,n^{\prime}}$ under a rotating wave approximation. This yields:
\begin{eqnarray}
G_{n,n^{\prime}}&\simeq\left(2\delta_{n^{\prime},n}+\delta_{n^{\prime},n+1}+\delta_{n^{\prime},n-1}\right)\sum_{k}\left|f\left(k\right)\right|^{2}\Re\sum_{m}\textrm{Tr}_{B}\left\{ b_{k,m}^{\dagger}b_{k,m}\rho_{B}\right\} \\P_{n,n^{\prime}}&\simeq\left(2\delta_{n^{\prime},n}+\delta_{n^{\prime},n+1}+\delta_{n^{\prime},n-1}\right)\sum_{k}\left|f\left(k\right)\right|^{2}\Re\sum_{m}\textrm{Tr}_{B}\left\{ b_{k,m}b_{k,m}^{\dagger}\rho_{B}\right\} 
\end{eqnarray}
where one can see that for both, gain and loss contributions, the local term is always twice as large as the non-local ones. This relation between the local and the non-local part is what makes the fermionic version of the H-N model topologically trivial for arbitrary values of the dissipative parameters.
\end{widetext}

\bibliographystyle{apsrev4-2}
\bibliography{Bibliography}

\end{document}